\documentclass[aps,prx,reprint,showpacs,showkeys,noeprint,longbibliography]{revtex4-1} 
\usepackage{amsmath}
\usepackage{latexsym}
\usepackage{float}
\usepackage[utf8x]{inputenc}
\usepackage[T1]{fontenc}
 \usepackage{amssymb}
\usepackage{graphicx}
\usepackage{textcomp}
\usepackage{hyperref}
\usepackage{subfigure}
\usepackage{xcolor}
\usepackage{url}
\usepackage{booktabs,tabularx,dcolumn}

\def\ba{\begin{eqnarray}}
\def\ea{\end{eqnarray}}
\def\be{\begin{equation}}
\def\ee{\end{equation}}
\def\bm{\begin{math}}
\def\me{\end{math}}

\newcommand{\dummy}

\begin{document}
\title{Effects of Alignment Activity on the Collapse Kinetics of a Flexible Polymer}
\author{ Subhajit Paul$^1$} \email[]{subhajit.paul@itp.uni-leipzig.de}
\author {Suman Majumder$^1$}\email[]{suman.majumder@itp.uni-leipzig.de}
\author {Subir K. Das$^2$}\email[]{das@jncasr.ac.in}
\author{Wolfhard Janke$^1$}\email[]{wolfhard.janke@itp.uni-leipzig.de}
\affiliation{$^1$Institut  f\"{u}r Theoretische Physik, Universit\"{a}t Leipzig, IPF 231101, 04081 Leipzig, Germany\\
	$^2$Jawaharlal Nehru Centre for Advanced Scientific Research, Jakkur P.O., Bangalore 560064, India}

\date{\today}

\begin{abstract}
Dynamics of various biological filaments can be understood within the framework of active polymer models. 
Here we consider a bead-spring model for a flexible polymer chain in which the active interaction among 
the beads is introduced via an alignment rule adapted from the Vicsek model.  Following a quench from the high-temperature coil phase to a low-temperature state point, 
we study the  coarsening kinetics via molecular dynamics (MD) simulations using the  Langevin thermostat. 
For the passive polymer case the low-temperature equilibrium state is a compact globule. 
Results from our MD simulations reveal that though the globular state is also  the typical final state in the active case, 
	the nonequilibrium pathways to arrive at such a state differ from the passive picture
	due to the alignment interaction among the beads.  
	We notice that deviations from the intermediate ``pearl-necklace''-like arrangement, that is
	observed in the passive case,
and the formation of more elongated dumbbell-like structures increase with increasing activity. 
Furthermore, it appears that while a small active force on the beads certainly makes the coarsening process much faster, 
there exists nonmonotonic dependence of the collapse time on the strength of active interaction.  
We quantify these observations by comparing the scaling laws for the collapse time and growth of pearls with the passive case.
\end{abstract}

\maketitle
\section{ Introduction}
\par 
Active matter systems have received much attention in the past few decades \cite{ramasw1, cates1, elgeti1, shaeb}. Compared to passive matter, the 
most distinguishing feature is their ability to self-propel and perform work either by consuming their internal energy or by drawing energy from the environment: Active matter is inherently out-of-equilibrium. As a consequence, active macro- and micro-organisms can form exotic large-scale patterns within which highly organized and coherent motion of the constituents is visible. For instance, certain active or dynamic interactions alone can lead to flocking behavior \cite{fily,vicsek}, which is analogous to vapor-liquid transitions in passive systems \cite{fily,redner,skdas14,trefz16,das1,paul1,vicsek}. Understanding the dynamics of active matter is hence of relevance for a wide variety of fields ranging from biology to society.

While the phenomenology of active {\em particle\/} systems is well explored \cite{ramasw1, cates1, elgeti1, shaeb,vicsek,toner,chate2,tailleur,chate1,jiang,hagen,mcand,mishra1,menzel,fily,farrell,redner,trefz16,das1,desei,paul1,skdas14}, active {\em polymers\/} have been considered only more recently \cite{winkler2020rev,holder,kaiser,ravi,sarkar,duman,bianco,paul_soft20,ramirez,biswas,daiki}. Here, interest primarily focused on the dynamics of the collective behavior \cite{winkler2020rev,holder,duman}, but to date still relatively little is known on the kinetics of the collapse transition of a single active polymer and the emerging structures during the associated coarsening process. For passive polymers, on the other hand, along with the equilibrium aspects of this transition \cite{stockmayer,gennes,sun,doi}, the 
latter nonequilibrium phenomena \cite{byrne,halperin, montesi,guo,majumder1,bunin,christiansen, majumder3, majumder4} have been studied extensively 
for many
years. By exploiting the analogy to particle and spin systems \cite{bray,puri}, various scaling properties of nonequilibrium statistical physics have been established with a fair degree of quantitative accuracy (for a recent review, see Ref.~\cite{majumder4}).
Typically these studies were performed by quenching a polymer from its high temperature extended coil state to a low temperature, and then monitoring its relaxation towards the compact globular state. This relaxation occurs 
via  an interesting pathway which is also of relevance for the dynamics of protein folding \cite{camacho, reddy} or the structural organization of chromatin \cite{shi2018}. Thus, understanding the nonequilibrium pathway as well as various scaling laws associated with this coil-globule transition has wide spread importance.
\par 
Amongst a few phenomenological descriptions of the collapse kinetics, the
``pearl-necklace'' picture of Halperin and Goldbart \cite{halperin} is quite well accepted and has been observed in coarse-grained as well as in all-atom
polymer models \cite{majumder1,christiansen,byrne,guo,montesi,majumder2}. 
The collapse starts with the formation
of a few small clusters (‘pearls’) along the chain which grow by withdrawing
monomers from the chain connecting them, eventually making the chain
stiffer. Then, in the subsequent coarsening stage, these clusters coalesce with each
other to form larger ones until a single cluster forms. In the final stage,
the monomers within this cluster rearrange to form a compact globule,
compatible with a minimum surface energy \cite{schnabel}. Two main interests in this problem
are to elucidate the scaling of collapse time and the growth of the average size of the clusters which is a measure of the relevant length scale of the coarsening process.
 \par 
\par
Whereas these aspects of the nonequilibrium kinetics are quite well understood for lattice and off-lattice models of passive polymers \cite{majumder1, christiansen, majumder3, majumder4,byrne,guo,montesi,bunin}, for active polymers there have been only recent computational efforts within Langevin or Brownian dynamics frameworks, with added self-propulsion through an extra active force term \cite{holder, duman, ravi, kaiser, bianco, 
	sarkar,paul_soft20}. For instance, Bianco et al. \cite{bianco} have observed in Brownian dynamics simulations an activity induced collapse transition of a single polymer.
In experiments also, filaments with active elements are being realized, e.g., 
by joining chemically synthesized artificial colloid or Janus particles via DNAs
\cite{jiang, biswas, daiki, ramirez}. 
External magnetic or electric fields can make them motile by providing phoretic motions of different types \cite{biswas, daiki}.
\par 
Inspired by these developments, here we consider a model of an active polymer in space 
dimension $d=3$. In this model, to each bead of a flexible bead-spring polymer chain \cite{milchev, majumder1, majumder3}, an ``active'' element 
is added. The activity is introduced via the well known Vicsek model \cite{vicsek,das1,paul1}. According to the general framework of this model, the beads intend to align their velocities with the average directions of their neighbors. In absence of the activity, the polymer chain is referred to as a ``passive'' one. The objective of this study is to quantify the 
effect of the activity on the kinetics of the coil-globule transition by comparing the results with those from the passive limit \cite{majumder1,christiansen, majumder3,majumder4}.
\par
The rest of the paper is organized as follows. 
In Sec.\ \ref{model} we will discuss the model and methods. 
Section\ \ref{result} contains the results. Finally we conclude the paper 
in Sec.\ \ref{conclusion}.

\section{ Model and Methods}\label{model}
\par 
We consider a coarse-grained bead-spring model for a single flexible polymer chain in which  
 $N$  monomers are connected in a linear fashion. The beads are made active by introducing Vicsek-like 
alignment interaction.
First we quantitatively discuss the passive interaction potentials acting 
among the beads.
The monomer-monomer bonded interaction is modelled via the standard finitely extensible non-linear elastic (FENE) potential \cite{majumder3, majumder4, milchev, majumder1}
\begin{equation}
V_{\rm{FENE}}(r) = - \frac{K}{2}R^2 {\rm{ln}} \bigg[ 1- \bigg(\frac{r-r_0}{R}\bigg)^2\bigg],
\end{equation}
where $r_0~(=0.7)$ is the equilibrium bond distance. The value of the spring constant $K$ has been set to $40$ and $R$, which measures the maximum deviation from $r_0$, to $0.3$.
\par 
The interaction between two non-bonded monomers at a distance $r$ apart is modeled via the standard Lennard-Jones (LJ) potential \cite{majumder1, majumder3,das1}
\begin{equation}
 V_{\rm{LJ}}(r) = 4\epsilon \bigg[\bigg(\frac{\sigma}{r}\bigg)^{12}- \bigg(\frac{\sigma}{r}\bigg)^6\bigg],
\end{equation}
with the interaction strength $\epsilon =1$. The diameter $\sigma$ of the monomers is chosen as $\sigma= r_0/2^{1/6}$.
The repulsive part in $V_{\rm{LJ}}$ takes care of the volume exclusion of 
the monomers. The attractive part will ensure the coil-globule transition 
for the polymer. A purely repulsive potential (e.g., of Weeks-Chandler-Andersen form \cite{wca_71}) 
cannot provide us a globular state.
For advantages in the numerical simulations the LJ potential has been truncated and shifted at $r_c=2.5\sigma$ such that the non-bonded pairwise 
interaction becomes \cite{frenkel}
\begin{equation}
  V_{\rm{NB}}(r)=
\begin{cases}
  V_{\rm{LJ}}(r)-V_{\rm{LJ}}(r_c) -(r-r_c)\frac{dV_{\rm{LJ}}}{dr}\Big|_{r=r_c} ~~ r<r_c \,,\\
 0 ~~~~~~~~~ \text{otherwise}\,.
   \end{cases}
\end{equation}
This modified potential should provide the same qualitative behavior as $V_{\rm{LJ}}$ and is also continuous and differentiable at $r=r_c$.\\

\par
The dynamics of this  polymer model has been studied via molecular dynamics (MD) simulations using the standard velocity-Verlet integration scheme \cite{frenkel}.  
To keep the temperature for the polymer constant at the quenched value, we employ the Langevin thermostat. 
Thus, for each bead we solve the equation
\begin{equation}\label{langevin}
 m_i \ddot{\vec{r}}_i = - \vec{\nabla} U_i - \gamma \dot{\vec{r}}_i + \sqrt{2 \gamma k_B T } \vec{\Lambda}_i(t),
\end{equation}
where $m_i=m$, the mass, is unity for all the beads,  $\gamma$ is the drag coefficient, for which we choose $\gamma=1$, $k_B$ is the Boltzmann constant, value of which 
is set to unity, and $T$ is the quench temperature, measured in units of $\epsilon/k_B$. The total energy $U_i$  contains contributions from LJ ($V_{\rm{LJ}}$) and FENE ($V_{\rm{FENE}}$) potentials.
 In Eq.~(\ref{langevin}) $\vec{\Lambda}(t)$ denotes Gaussian random noise with zero mean and unit variance. This is  Delta correlated over
space and time as 
\begin{equation}
 \langle\Lambda_{i\mu}(t)\Lambda_{j\nu}(t')\rangle = \delta_{ij}\delta_{\mu \nu} \delta(t,t'),
\end{equation}
where $i,j$ are the particle indices and $\mu$, $\nu$ correspond to 
the cartesian coordinates. Time $t$ in our simulations has been measured in 
units of $\tau_0 = \sqrt{m \sigma^2/\epsilon}$. We have chosen the time 
step of integration $\delta t$ as $5 \times 10^{-4}$ in units of $\tau_0$. Solution of Eq.~(\ref{langevin}) as a function of time provides the dynamics of the passive polymer.

\par
Now we describe how the activity is incorporated in the 
Vicsek manner \cite{vicsek, das1, skdas14, trefz16,paul1}.
At each MD step, the velocity for the $i$-th bead ($\vec{v}_i^{\,\rm{pas}}(t+\delta t)$), obtained from Eq.~(\ref{langevin}), is modified by an additional force ($\vec{f}_i$)
\begin{equation}\label{active_force}
 \vec{f}_i = f_A \hat{v}_i^{\rm{avg}} ,
\end{equation}
where $f_A$ is a constant that measures the strength of the active force. 
 For the passive polymer $f_A=0$. 
The unit vector $\hat{v}_i^{\rm{avg}}$ represents the average direction of the velocities of the neighboring 
beads within a certain radius $r_v$ around the bead $i$ and is calculated 
as 
\begin{equation}\label{neighbor}
 \hat{v}_i^{\rm{avg}} = \frac{\Big(\sum_j \vec{v}_j\Big)_{r_v}}{\big|\Big(\sum_j \vec{v}_j\Big)_{r_v}\big|}\,,
\end{equation}
 in which we choose $r_v=r_c$. Due to the active force the velocity of the $i$-th bead gets modified to $\vec{v}_i^{\,*}$ as
 \begin{equation}\label{resultant}
  \vec{v}_i^{{\,*}}(t+\delta t) = \vec{v}_i^{\,\rm{pas}}(t+\delta t) + \frac{\vec{f}_i}{m_i} \delta t\,.
 \end{equation}
When applied in this manner, the active force would change the direction as well as the magnitude of the velocity, which may alter the temperature 
felt by the polymer. As the role of Vicsek activity \cite{vicsek} is only to change  the direction of the velocities of the beads, we need to rescale the magnitude of $\vec{v}_i^{\,*}$ to its passive value leading to the final velocity  \cite{das1}
\begin{equation}
 \vec{v}_i^{{f}}(t+\delta t) = \mid \vec{v}_i^{\,\rm{pas}}(t+\delta t) \mid\hat{n}_i\,,
\end{equation}
where $\hat{n}_i$ is the unit vector of $\vec{v}_i^{{\,*}}$ emerging from  Eq.~(\ref{resultant}).
Increase of the strength of $\vec{f}_i$, i.e., the activity, by varying the value of $f_A$, as defined in  Eq.~(\ref{active_force}), is expected to help the velocities of the beads to align themselves more rapidly. 
\par
For both passive and active cases, the initial configurations have been prepared at high temperature where the polymer is in an extended coil state. 
For studying the nonequilibrium dynamics the polymer is quenched to a temperature $T_f=0.5$, which is below the coil-globule transition temperature ($T_{\theta}$) for the passive case \cite{majumder3}. We studied polymer chains with the number of beads $N$ varying over a wide range between $32$ and $512$. In   each of the cases, all the presented data are averages  over $500$ independent initial realizations (initial conditions and thermal noise).

\section{ Results}\label{result}
\begin{figure*}[t!]
	\centering
	\includegraphics*[width=0.96\textwidth]{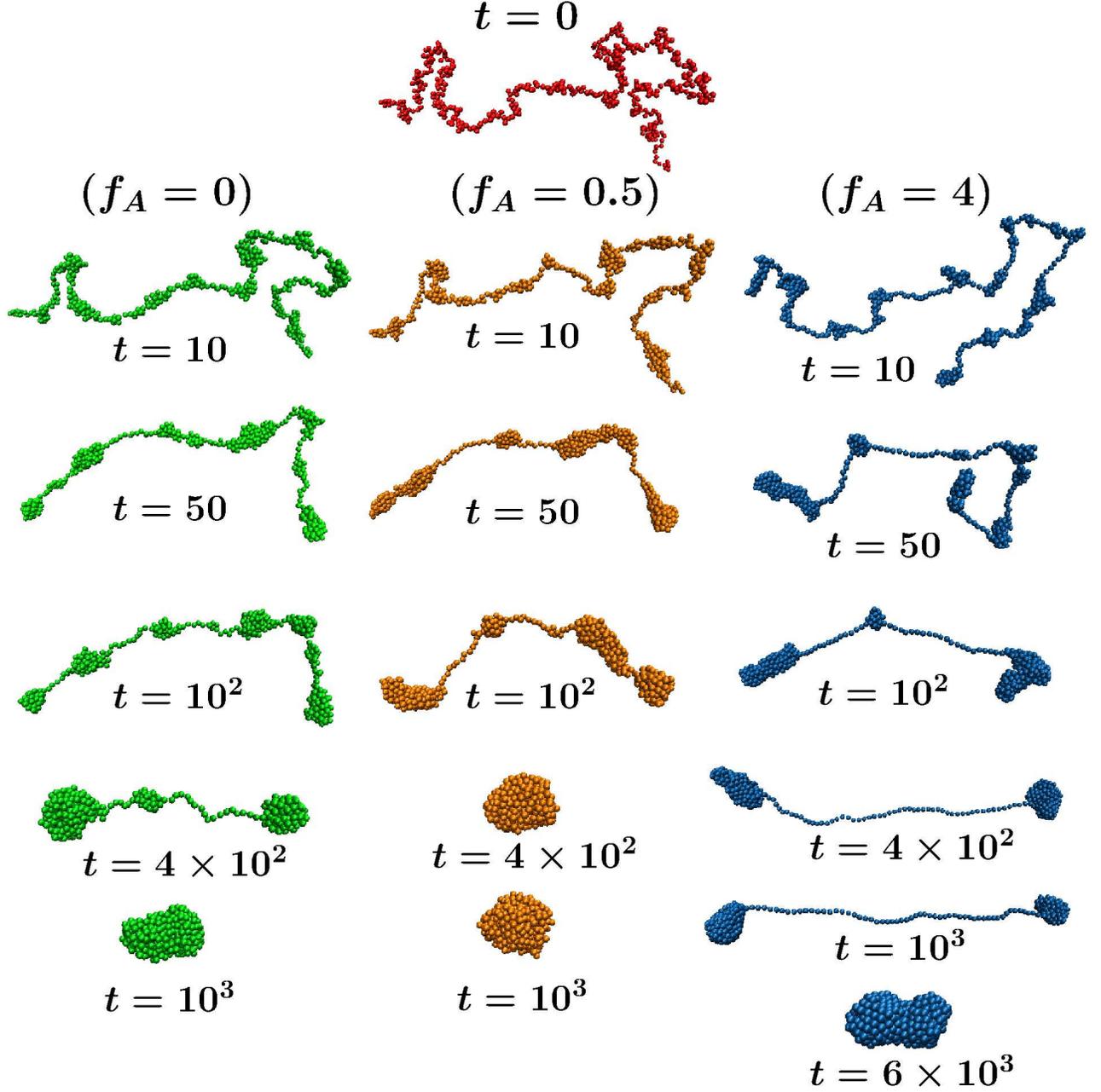}
	\caption{\label{snap} Snapshots obtained at different times during the evolution, are shown for  
		the collapse of a polymer with $N=512$.  We have included results for 
the passive ($f_A=0$) as well as the active cases ($f_A=0.5$ and $4$). 
		The times are mentioned below each of the conformations.}
\end{figure*}
In Fig.~\ref{snap} we show snapshots that were recorded during the evolution of a polymer chain towards its collapsed states. There we have included results from the passive ($f_A=0$) as well as active cases ($f_A=0.5$ and $4$). In all the
cases we have started with the same coil conformation shown at the top ($t=0$). With increasing time, the sequence of events for $f_A=0$ and $0.5$ are consistent with the ``pearl-necklace'' model of Halperin and Goldbart \cite{halperin}. It appears that for $f_A=0.5$ coarsening occurs faster. Whereas for the passive polymer there exist a few 
clusters at $t=400$, the chain with $f_A=0.5$ has already fully collapsed.
\par 
For the highest activity $f_A=4$, the structures at intermediate times (say, $t=50$ or $100$) look somewhat different than those for smaller $f_A$. For $f_A=4$, even though the morphology has similarity with the ``pearl-necklace'' picture, the conformations appear stiffer and elongated. There exists another interesting feature, at late time: a dumbbell-like conformation persists for a rather long time, before the clusters at the two ends finally coalesce to form a single globule. Also, this globule has an elongated or sausage-like structure. This is unlike other cases for which the final structure is more spherical. Here we should mention that the overall relaxation time for arriving at the sausage structure, for $f_A=4$, has a rather broad distribution, as will be shown later.

\begin{figure}[t!]
	\centering
	\includegraphics*[width=0.5\textwidth]{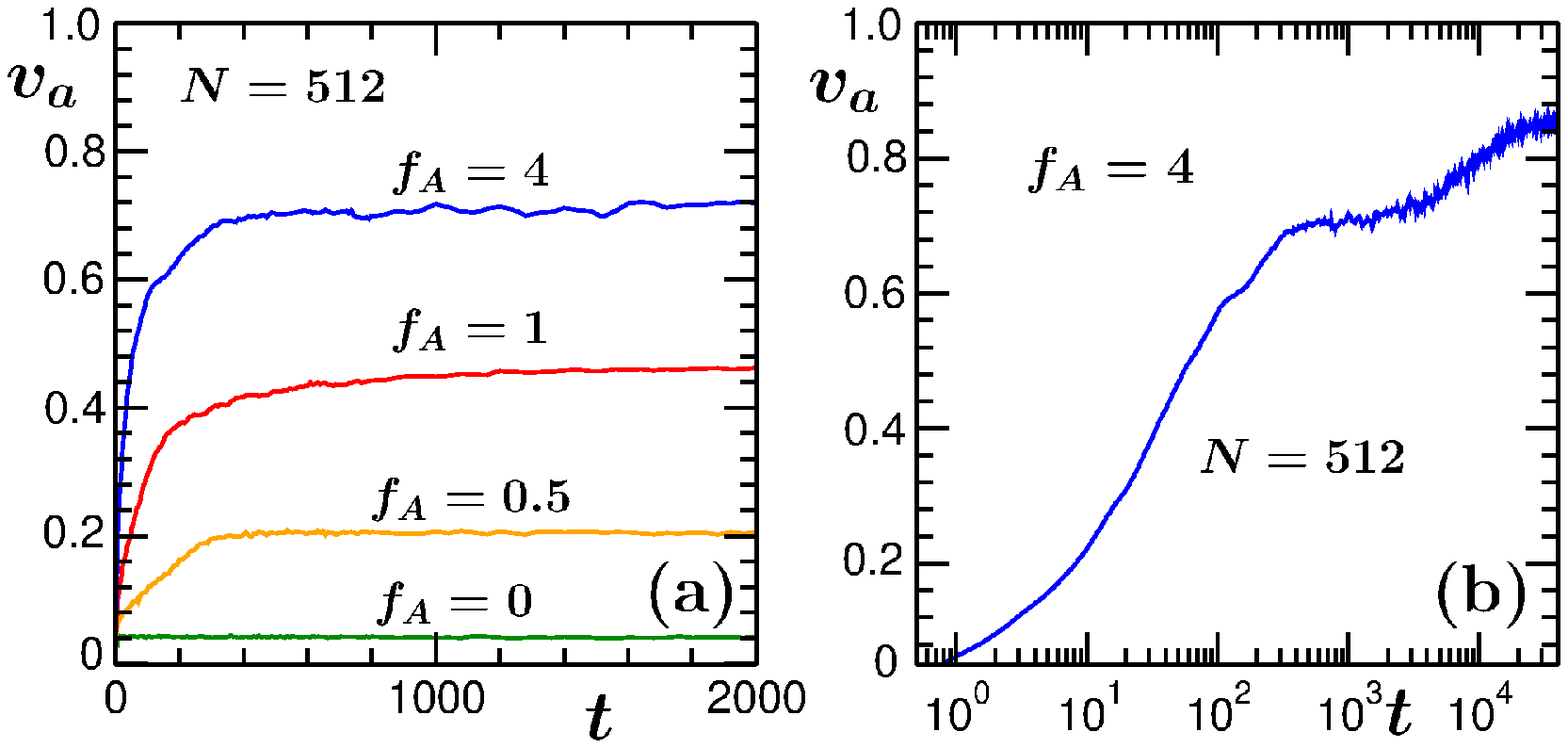}
	\caption{\label{opm} (a) Plot of $v_a$, the order parameter, as a function of time for different values of $f_A$, with $N=512$. (b) The same plot
for $f_A=4$ as in (a) over a much longer time period on a semi-log scale.}
\end{figure}
\par
Before going for the more quantitative analysis to compare the relaxation times among the different $f_A$ values, we will first discuss the main effect of the increasing activity among the beads. 
For this, following Ref.~\cite{vicsek}, we calculate the velocity order parameter, 
\begin{equation}\label{opv}
v_a = \frac{\mid\sum\limits_{i=1}^{N} \vec{v}_i\mid}{\sum\limits_{i=1}^{N} \mid\vec{v}_i\mid}\,,
\end{equation}
where $\vec{v}_i$ denotes the velocity of the $i$-th bead. In Fig.~\ref{opm}(a) we show the time dependence of $v_a$ for different values of $f_A$ using polymers of length $N = 512$. Clearly, for the passive case $v_a$ remains at a constant value close to zero throughout. For nonzero $f_A$, $v_a$ saturates at values that increase  with $f_A$. When $f_A$ is very high, say, for $f_A=4$, $v_a$ exhibits a rapid increase at early time. The data for this case is shown in Fig.~\ref{opm}(b) on a semi-log scale, over a longer range. Clearly a two-step saturation is visible. 
This is consistent with what one can read off from Fig.~\ref{snap}. There, for $f_A=4$, the dumbbell starts forming at $t=300$, matching with the arrival of the first plateau in Fig.~\ref{opm}(b). Following this, as the two clusters 
remain separated, the value of $v_a$ does not alter much. Much later, when the two clusters come closer, assisted by thermal fluctuations and the attraction among the monomers, $v_a$ starts increasing again. This fact, perhaps, indicates that the velocity alignments among the beads start earlier than the coarsening in the density field. More quantitative information on this will be provided later.
\par
The rest of the section we divide into two parts. In the first subsection 
we discuss results related to the relaxation times. The second one is devoted to the results on growth kinetics of clusters.
\begin{figure}[b!]
	\centering
	\includegraphics*[width=0.43\textwidth]{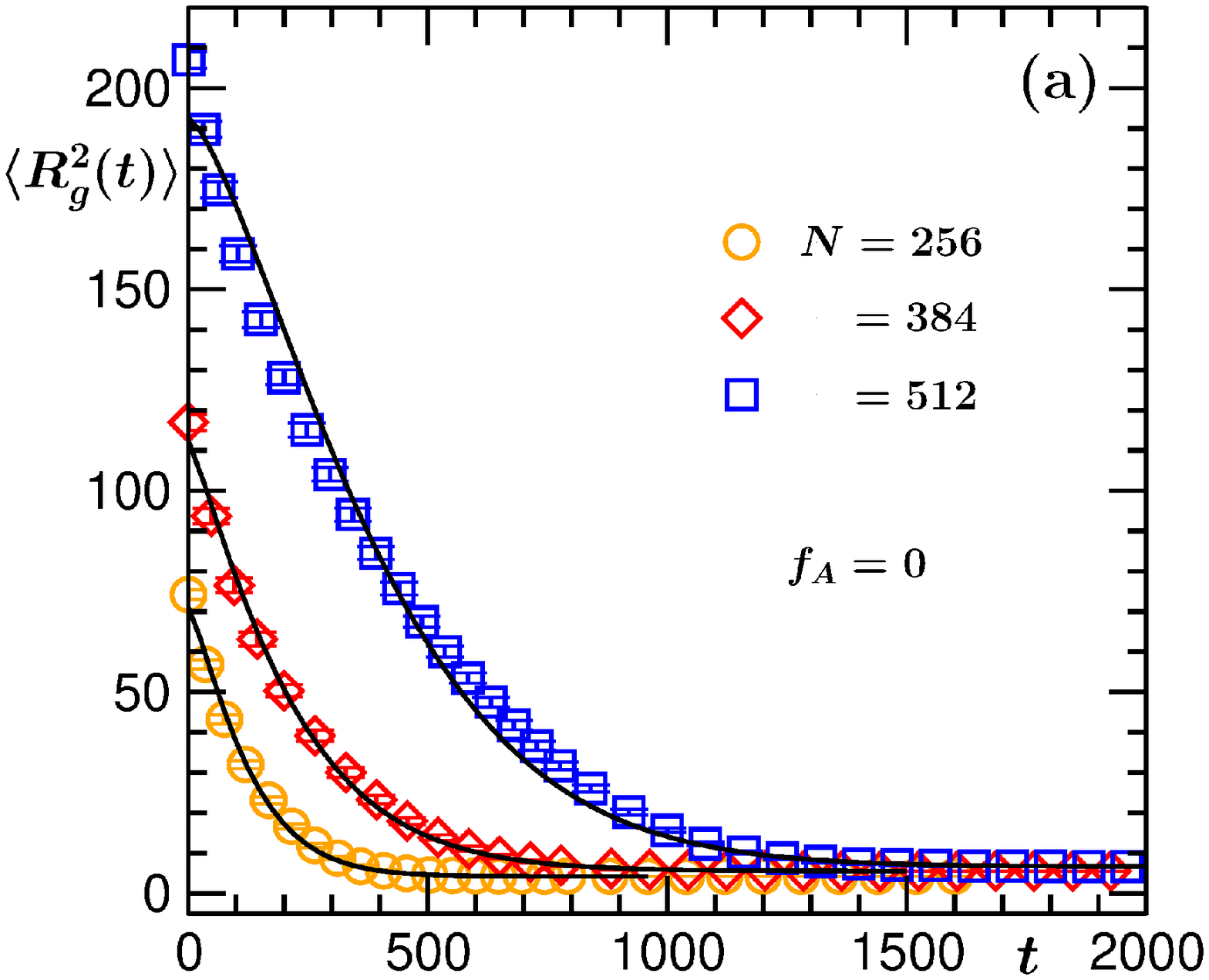}
	\vskip 0.4cm
	\includegraphics*[width=0.41\textwidth]{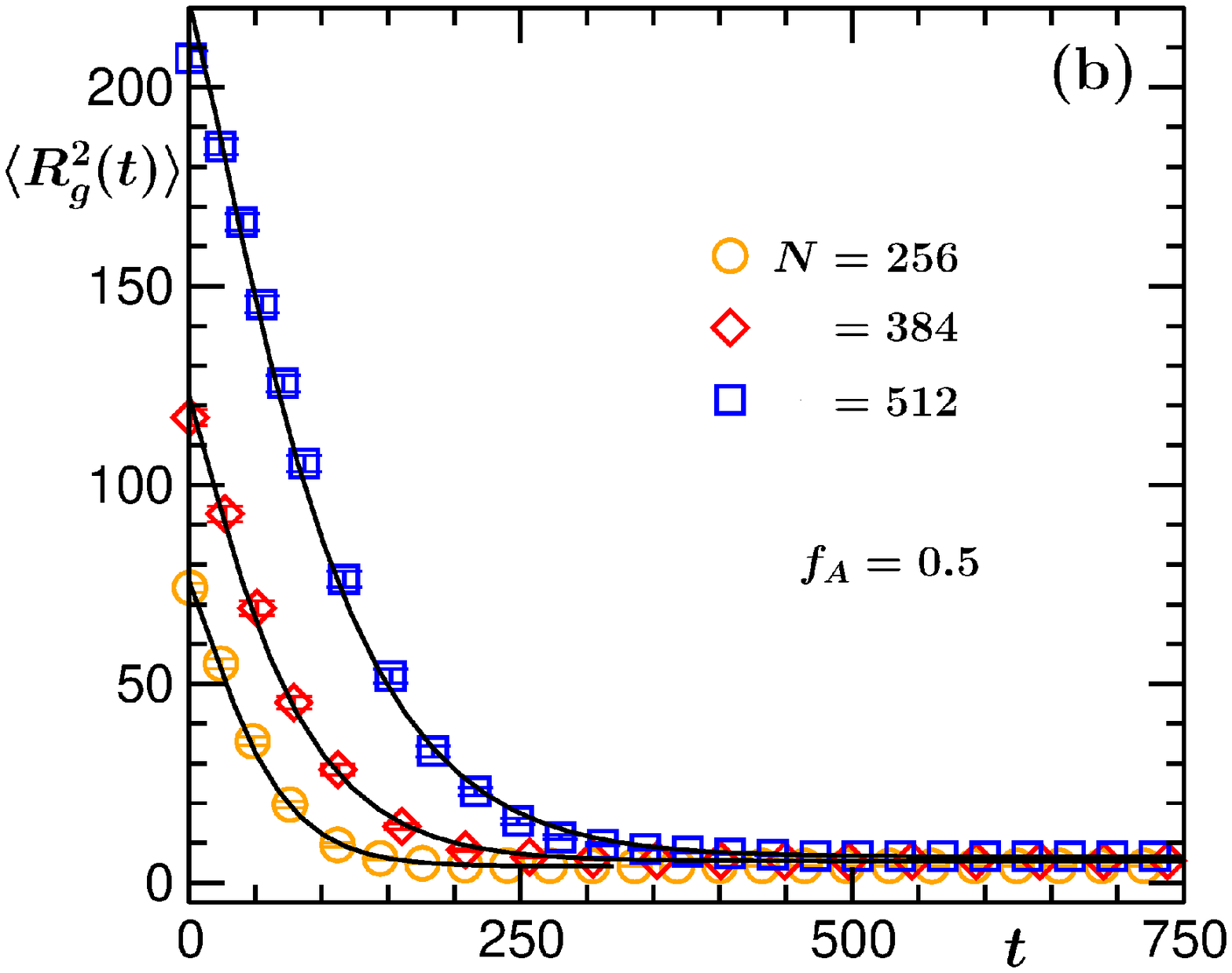}
	\caption{\label{rgsq_decay} (a) Average squared radius of gyration, $\langle R_g^2 \rangle$, versus $t$, for three different polymer lengths, as mentioned in the figure, for $f_A=0$. Notice the delayed decay with the increase of the chain length. The solid lines show fits using the functional form \eqref{rgsq_fit}. (b) Same as (a) for $f_A=0.5$.}
\end{figure}
\subsection{Relaxation times and associated scaling}
To quantify the relaxation time for the coil-globule transitions it is important to have ideas about the radius of gyration of a polymer:
\begin{equation}\label{rgsq_def}
R_g^2 = \frac{1}{N}\sum_{i=1}^{N} (\vec{r}_{\rm{cm}} - \vec{r}_i)^2\,.
\end{equation}
Here $\vec{r}_{\rm{cm}}$ is the center-of-mass of the polymer, given by
\begin{equation}\label{com_def}
\vec{r}_{\rm{cm}}=\frac{1}{N}\sum_{i=1}^{N} \vec{r}_i\,.
\end{equation}
In Fig.~\ref{rgsq_decay}(a) we show plots of the average squared radius of gyration, $\langle R_g^2 \rangle$, versus time, for different chain lengths $N$, by fixing $f_A$ to zero. The angular brackets $\langle ... \rangle$ represent an average over different initial configurations and time evolutions.
The decay to the final or steady state value appears to occur later for longer chain lengths. In fact, as we will see later, it becomes slower as well.
We fit our data sets with the ansatz 
\begin{equation}\label{rgsq_fit}
\langle R_g^2(t) \rangle =b_0+b_1 \exp[-(t/\tau_c)^\beta]\,,
\end{equation}
where $b_0$ is the value of $\langle R_g^2 (t \to \infty) \rangle$ in the collapsed state of the polymer, $\tau_c$ is the relaxation time for approaching the latter state, and $b_1$ and $\beta$ are other adjustable or fitting parameters,
with $b_1$ being related to the value of $\langle R_g^2 \rangle$ at $t=0$. The solid lines in this figure represent best fits of the ansatz \eqref{rgsq_fit} 
to the simulation data. Figure\ \ref{rgsq_decay}(b) contains results for $f_A=0.5$. In each of these frames we have included analysis for three different chain lengths. For both the values of $f_A$, a common trend is observed: $\tau_c$ increases with the increase of $N$.
\par 
The values, as well as the order of the errors, of the fitting parameters, for $f_A=0$, obtained by using Jackknife resampling techniques \cite{efron}, are quoted in Table~\ref{tab1}. There we have included the values 
of the parameters for a few more chain lengths, in addition to those shown in Fig.~\ref{rgsq_decay}(a). For nonzero $f_A$ also, the relaxation is well described by the exponential form (\ref{rgsq_fit}). The corresponding values of the fitting parameters for $f_A=0.5$ are compiled in Table~\ref{tab2}.

\begin{table}[b!]
	\caption{ Parameters $b_0$, $b_1$, $\tau_c$, and $\beta$
	for passive polymers of lengths $N$, obtained by fitting the time dependence of the average squared radius of gyration  $\langle R_g^2 \rangle$ with the ansatz \eqref{rgsq_fit}.}\label{tab1}
\centering
\begin{tabular}{|c|c|c|c|c|}
\hline
~$N$~~&~~$b_0$~~&~~$b_1$~~&~~$\tau_c$~~&~~$\beta$~~\\
\hline
~~32~~&~~1.15(1)~~&~~5.5(1)~~&~~9.8(2)~~&~~1.13(3)~~\\
~~64~~&~~1.767(9)~~&~~13.6(3)~~&~~22.5(4)~~&~~1.21(3)~~\\
~~128~~&~~2.729(8)~~&~~30.7(6)~~&~~56(1)~~&~~1.26(4)~~\\
~~192~~&~~3.529(8)~~&~~45.7(8)~~&~~89(2)~~&~~1.25(2)~~\\
~~256~~&~~4.241(2)~~&~~67(1)~~&~~134(3)~~&~~1.24(3)~~\\
~~384~~&~~5.501(6)~~&~~107(2)~~&~~227(6)~~&~~1.17(5)~~\\
~~512~~&~~6.614(5)~~&~~186(2)~~&~~437(6)~~&~~1.40(5)~~\\
\hline
\end{tabular}
\end{table}

\begin{table}[t!]
	\caption{Same as Table~\ref{tab1} for the active case with $f_A=0.5$.}\label{tab2}
\centering
\begin{tabular}{|c|c|c|c|c|}
\hline
~$N$~~&~~$b_0$~~&~~$b_1$~~&~~$\tau_c$~~&~~$\beta$~~\\
\hline
~~32~~&~~1.121(9)~~&~~5.5(2)~~&~~8.7(2)~~&~~1.19(3)~~\\
~~64~~&~~1.735(6)~~&~~14.3(3)~~&~~16.6(4)~~&~~1.24(2)~~\\
~~128~~&~~2.698(7)~~&~~33.3(6)~~&~~29.9(7)~~&~1.23(3)\\
~~192~~&~~3.499(8)~~&~~49.7(9)~~&~~40(1)~~&~~1.20(3)~~\\
~~256~~&~~4.214(3)~~&~~72(2)~~&~~53(1)~~&~~1.24(2)~~\\
~~384~~&~~5.476(2)~~&~~117(2)~~&~~72(2)~~&~~1.15(2)~~\\
~~512~~&~~6.595(1)~~&~~215(2)~~&~~101(2)~~&~~1.21(2)~~\\
\hline
\end{tabular}
\end{table}
\par
\begin{figure}[t!]
	\centering
	\includegraphics*[width=0.46\textwidth]{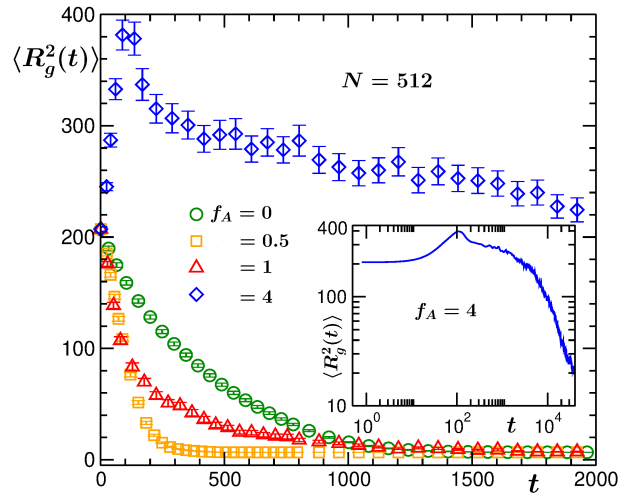}
	\caption{\label{rg_fa} Plots of $\langle R_g^2 \rangle$ vs time, for a polymer of length $N=512$ and different values of 
	activities. The inset shows the data for $f_A=4$ on a log-log scale over a much longer time.}
\end{figure}

 
To obtain an idea about the trend of relaxation with the variation of $f_A$, in Fig.~\ref{rg_fa} we plot $ \langle R_g^2 \rangle$ as a function of time, for several values of $f_A$, by fixing the chain length at $N=512$. Unless otherwise mentioned, from now onwards, we will use  this value of $N$. As previously pointed out, there exists nonmonotonicity  in the dependence of $\tau_c$ on $f_A$. Compared to the $f_A=0.5$ case, for which the globule forms earlier than in the passive case, the process becomes slower when the activity becomes much higher. Figure~\ref{rg_fa} reveals another interesting fact: it can be convincingly concluded that the chain gets more elongated for $f_A=4$ in the early period of evolution. Thus, it is not possible to fit the entire range of the data with the ansatz \eqref{rgsq_fit} that assumes a monotonic decay.
\par 
In Eq.~(\ref{rgsq_fit}) $\beta=1$ implies a simple exponential decay. From fitting, however, in all the cases, we obtain values slightly higher than unity. Values of $\tau_c$ by fixing $\beta=1$ do not differ much from those quoted in Tables~\ref{tab1} and \ref{tab2} for which $\beta$ was chosen as a fitting parameter.
\par  
Alternatively, the collapse or relaxation time $\tau_q$ can also be defined as the time at which $R_g^2$, starting from its value at $t=0$,  reaches to a fraction $q$ of its total decay \footnote{Note that this definition slightly differs from the one used in Ref.\ \cite{majumder3}. The parameter $p$ used there relates to $q$ as $q=1-p$.}.  Considering the offset value of $R_g^2$ at $t \to \infty$, $\tau_q$  follows from \cite{majumder3} 
\begin{equation}\label{relax_time}
R_g^2 (t=\tau_q) = q \Delta R_g^2 + R_g^2(t \to \infty)\,,
\end{equation}
where $\Delta R_g^2 \equiv R_g^2 (t=0) - R_g^2 (t \to \infty)$ is the total decay of the gyration radius.  Since the decay of $R_g^2$ has been shown to follow a nearly exponential behavior, we choose $q=1/e$. Such a choice is motivated by Eq.~(\ref{rgsq_fit}) using $\beta=1$ (in this case, $\tau_q = \tau_c$). Though generic and employed recently for passive polymers \cite{majumder3,majumder4}, the definition (\ref{relax_time}) becomes questionable when $R_g^2$ does not decay monotonically with time. Since we observed a nonmonotonic behavior for $f_A=4$, this calls for a more direct method in such a situation.
\par 
 One such method for extracting the relaxation time is to consider the time when the polymer first forms a single cluster after the completion of the ``pearl-necklace'' stage. For this in the following we use the notation $\tau_{\rm{cl}}$. Estimation of $\tau_{\rm{cl}}$ requires the identification of clusters along the chain. For this purpose, we calculate the number of neighboring monomers around a bead, say the $i$-th one, within a cut-off distance $r_{\rm{cut}}$, as 
\begin{equation}
n_i=\sum_{j=1}^{N} \Theta (r_{\rm{cut}}-r_{ij})\,.
\end{equation}
Here $r_{ij}$ is the distance between the $i$-th and the $j$-th beads, and for  $r_{\rm{cut}}$ we uniformly use $2.5\sigma$. If $n_i$ is larger than a pre-set number, say, $n_{\rm{min}}$, a cluster is identified around the bead $i$ with $n_i$ monomers in it. Here we have chosen $n_{\rm{min}}=15$.  Using merely the above criterion would lead to an overestimation of the actual number of discrete clusters because clusters identified at neighboring beads are  typically part of one and the same  bigger cluster. Thus in a further step, we get rid of this overestimation by using a Venn diagram approach as described in Ref.~\cite{majumder4}, and obtained the correct number of distinct clusters $n_c(t)$ and the number of monomers within each cluster (i.e., its mass) $m_k$.  The time $\tau_{\rm{cl}}$ is obtained when $n_c(t)=1$ for the first time. The time dependence of $n_c(t)$ for different activities will be discussed in the next subsection.
\par
\begin{figure}[t!]
	\centering
	\includegraphics*[width=0.45\textwidth]{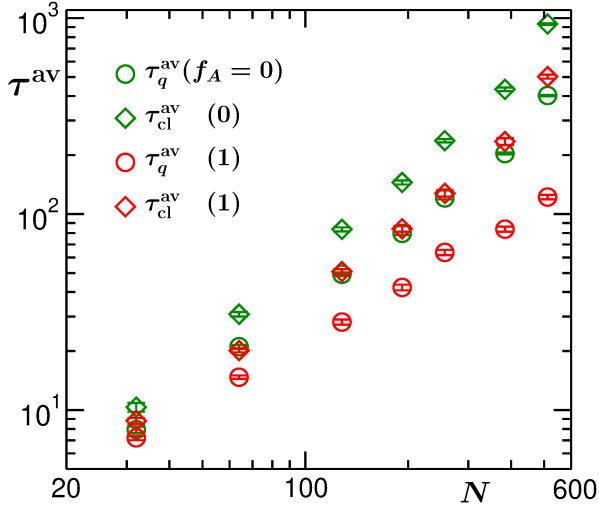}
	\caption{\label{tau_fitcl}  Plots of the average relaxation times $\tau^{\rm{av}}$, i.e., $\tau_q^{\rm{av}}$ (estimated from the decay of $R_g^2$), and  $\tau_{\rm{cl}}^{\rm{av}}$ (estimated from the direct method)
		versus the polymer chain length $N$, for $f_A=0$ and $1$.}
\end{figure}

 
In Fig.~\ref{tau_fitcl} we show the variation of both the relaxation times, i.e., $\tau_q^{\rm{av}} (= \langle \tau_q \rangle)$ and $\tau_{\rm{cl}}^{\rm{av}} (= \langle \tau_{\rm{cl}} \rangle)$, estimated from the decay of $R_g^2$ [Eq.~(\ref{relax_time})] and the direct method, respectively, as a function of $N$, for the passive case as well as for $f_A=1$, on a log-log scale. Again $\langle ... \rangle$ denotes averaging over different initial conformations and time evolutions. For $f_A=0$, the data sets corresponding to $\tau_q^{\rm{av}}$ and $\tau_{\rm{cl}}^{\rm{av}}$ are reasonably parallel 
to each other, differing mainly by a multiplicative factor, suggesting nearly the same power-law exponent for both of them. We checked that this is true for $f_A=0.5$ as well. For $f_A=1$, while for smaller values of $N$ both $\tau_q^{\rm{av}}$ and $\tau_{\rm{cl}}^{\rm{av}}$ are almost proportional to each other similar to that for $f_A=0$, they seem to deviate for large $N$, especially when $N > 256$. 
This could be due to the fact that for large $N$ the decay of $R_g^2$ from single runs is subject to strong fluctuations up to a certain time. This is certainly not visible from our plot of $\langle R_g^2 \rangle$ (in Fig.~\ref{rg_fa}) which is an average quantity.   Such fluctuations eventually disappear at late times. Also this effect is likely to be stronger for higher values of $f_A$. We have checked that if one uses $q=0.1$ in Eq.~\eqref{relax_time}, instead of $q=1/e$, the resultant data of $\tau_q^{\rm{av}}$ as a function of $N$ looks almost parallel to the corresponding data for $\tau_{\rm{cl}}^{\rm{av}}$. To avoid such inconsistency due to the sensitivity on the choice of $q$ in Eq.~\eqref{relax_time}, for further analyses of the relaxation time we will only consider $\tau_{\rm{cl}}$, obtained from the direct method.
It is more generic, particularly because it applies to all $f_A$ values -- recall the nonmonotonicity in Fig.~\ref{rg_fa}. In this context, one may think that the well-formed single clusters could break up or fragment, as observed for ``active'' clusters made up of active colloidal particles \cite{ginot18}.  But in our polymer model, due to the presence of bond connectivity which always maintains an upper bound in distance between two successive beads, along with the attractive interaction among the monomers, such a break-up of the globule is very unlikely.

  \par 
  For a more quantitative understanding of the distribution of $\tau_{\rm{cl}}$, obtained from different initial realizations and time evolutions,  in addition to the average $\tau_{\rm{cl}}^{\rm{av}} = \langle \tau_{\rm{cl}} \rangle$, we have also looked at certain higher central moments, namely, the variance $\sigma^2=\langle (\tau_{\rm{cl}}-\tau_{\rm{cl}}^{\rm{av}})^2\rangle$, the skewness $\mu_3=\langle \big(\frac{\tau_{\rm{cl}}-\tau_{\rm{cl}}^{\rm{av}}}{\sigma}\big)^3\rangle$ and the kurtosis  $\mu_4=\langle \big(\frac{\tau_{\rm{cl}}-\tau_{\rm{cl}}^{\rm{av}}}{\sigma}\big)^4\rangle$. For any distribution, $\mu_3$ measures the asymmetry with respect to its mean, whereas $\mu_4$ is a measure of the  heaviness of the data points in its tail. Often one considers the excess kurtosis,
  \begin{equation}
  \tilde{\mu}_4 = \mu_4 -3\,,
  \end{equation}
  where $3$ is the value of the kurtosis for a Gaussian distribution. In Table~\ref{tab3} we present these statistical quantities for $\tau_{\rm{cl}}$. There one sees that for higher $f_A$ the values of $\mu_3$ and $\mu_4$ are much higher compared to those for $f_A=0$  and $0.5$.

\par
\begin{figure}[b!]
	\centering
	\includegraphics*[width=0.48\textwidth]{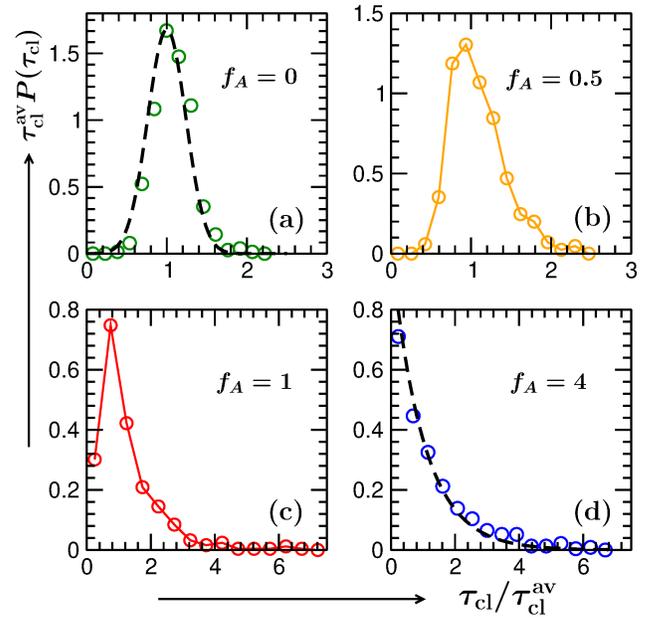}
	\caption{\label{tau_distri} Normalized distributions of $\tau_{\rm{cl}}$, for the passive as well as the active cases. The black dashed lines in (a) and (d) are the functions $f_1(\tau_{\rm{cl}})$ and $f_2(\tau_{\rm{cl}})$ given in Eqs.~(\ref{gauss_fit}) and (\ref{exp_fit}), respectively.  The solid lines in (b) and (c) are guides to the eye. All data sets are for $N=512$.}
\end{figure}

 
To have an idea about the distributions of $\tau_{\rm{cl}}$ with increasing activity,  in Figs.~\ref{tau_distri}(a)-(d) we show the normalized distributions $\tau_{\rm{cl}}^{\rm{av}}P(\tau_{\rm{cl}})$ versus $\tau_{\rm{cl}}/\tau_{\rm{cl}}^{\rm{av}}$ for different values of $f_A$, with $N=512$. When scaled  by the corresponding $\tau_{\rm{cl}}^{\rm{av}}$, the ranges for the abscissa variable become similar for all of them. Also the comparative understanding of the values of $\mu_3$ and $\mu_4$ for different values of $f_A$ becomes easier.  Being dimensionless, values of  these quantities do not get affected by such scaling. We see that only for the passive case in Fig.~\ref{tau_distri}(a) the distribution is a Gaussian for which a function with the form
\begin{equation}\label{gauss_fit}
f_1(\tau_{\rm{cl}})=\frac{\tau_{\rm{cl}}^{\rm{av}}}{\sqrt{2\pi\sigma^2}}~ {\exp\Big[-\frac{1}{2\sigma^2}\big({\tau_{\rm{cl}}-\tau_{\rm{cl}}^{\rm{av}}}\big)^2\Big]}\,,
\end{equation}
is plotted, where $\sigma^2$ is related to the width of the distribution. The corresponding values for $\tau_{\rm{cl}}^{\rm{av}}$ and $\sigma^2$ are taken from Table~\ref{tab3}.  Now for the plots in Figs.~\ref{tau_distri}(b)-(d) we see that  with the increase of the  activity the distributions deviate from being Gaussian and seem to exhibit a crossover to an exponential behavior. In fact, the dashed line in Fig.~\ref{tau_distri}(d) shows that the distribution for $f_A=4$ is compatible with an exponential function of the form 
\begin{equation}\label{exp_fit}
f_2(\tau_{\rm{cl}})=\exp\big(-\tau_{\rm{cl}}/\tau_{\rm{cl}}^{\rm{av}}\big)\,,
\end{equation} 
where $\tau_{\rm{cl}}^{\rm{av}}=6500$, as given in Table~\ref{tab3}. For a perfectly exponential decay one expects $\mu_3=2$ and $\tilde{\mu}_4=6$. While our empirically estimated value for $\mu_3$ agrees quite well, the excess kurtosis $\tilde{\mu}_4$ appears quite underestimated. We have checked that this is the general trend for low statistics such as with only $500$ events used here.

\begin{table}[t!]
	\caption{Various moments of the $\tau_{\rm{cl}}$ distributions for $N=512$ obtained from $500$ realizations, for the passive as well as the active cases.}\label{tab3}
	\centering
	\begin{tabular}{|c| c|c|c|c|}
		\hline
		\hskip 0.1cm
		~~$f_A$~~ &~~ $\tau_{\rm{cl}}^{\rm{av}}$~~ & ~~$\sqrt{\sigma^2}$~~ &~~ $\mu_3$~~ & ~~$\tilde{\mu}_4$~~    \\
		\hline
		~~$0$~~ &~~ 932.4~~ & ~~222.3~~ & ~~0.367~~ & ~~0.55~~   \\
		
		~~$0.5$~~ &~~ 209.8~~  & ~~69.65~~ & ~~0.987~~ &~~1.20~~\\
		
		~~$1$~~ & ~~502.6~~  & ~~ 484.9~~ & ~~2.364~~  &~~ 7.665~~ \\
		
		~~$4$~~ & ~~6500~~ & ~~ 7496.8~~ & ~~1.70~~ & ~~2.95~~\\
		\hline 
	\end{tabular}
\end{table}

\par
\begin{figure}[b!]
	\centering
	\includegraphics*[width=0.46\textwidth]{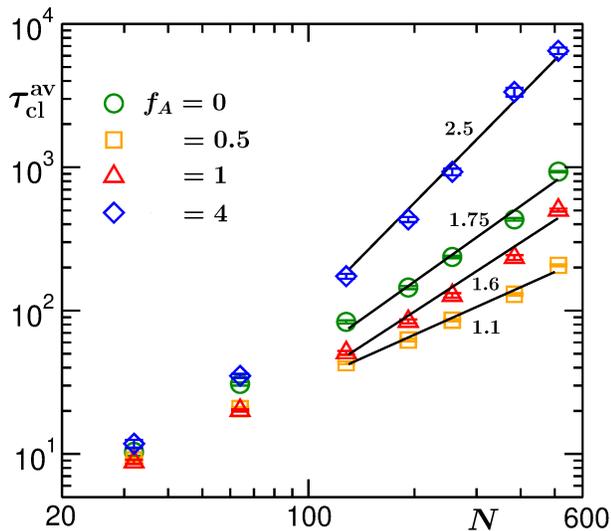}
	\caption{\label{tau_N0} Plots showing the scaling of the relaxation times, $\tau_{\rm{cl}}^{\rm{av}}$, as a function of $N$, 
		for different values of $f_A$. The solid lines represent various power laws.}
\end{figure}

 
In Fig.~\ref{tau_N0} we plot $\tau_{\rm{cl}}^{\rm{av}}$ versus $N$, for different values of $f_A$. For our shortest considered chain, i.e., $N=32$, the values of $\tau_{\rm{cl}}^{\rm{av}}$ for various $f_A$ are not too different from each other. Data for the passive case shows a scaling of the form 
\begin{equation}\label{tauN_sclng}
\tau_{\rm{cl}}^{\rm{av}} \sim N^{z},
\end{equation}
with $z \approx 1.75$.  This is in good agreement with previous reports \cite{majumder3, majumder4, klushin,  kikuchi}. Due to a lack of appropriate theoretical understanding in the nonequilibrium context, often $z$ is compared with the analogous exponent for equilibrium scaling, referred to as the Rouse scaling, for which $z=2$ \cite{rouse}.
\par 
For $f_A=0.5$, with $N \ge 64$, values of $\tau_{\rm{cl}}^{\rm{av}}$ are significantly smaller than the corresponding passive values. Here also, the data show a power-law scaling of the form (\ref{tauN_sclng}), however, 
with $z \approx 1.1$, which is much smaller than the exponent for the passive case. Interestingly,
for $f_A=1$ we see that the values of $\tau_{\rm{cl}}^{\rm{av}}$ are larger than those for $f_A=0.5$, for $N > 64$, but still smaller than the passive case. Also the errors in this case are slightly larger. This data look consistent with a power law with exponent $z \approx 1.6$. Now, for much higher activity, i.e., with $f_A=4$, the values of $\tau_{\rm{cl}}^{\rm{av}}$ for $N \ge 64$ are larger than for all the other $f_A$ cases and the exponent $z \approx 2.5$ is naturally much larger. As already observed from Table~\ref{tab3} for high activity the values for the mean as well as the standard deviation are much larger. Again, this is an interesting observation, indicating that a competition between activity strength and thermal fluctuation is important.

\par
Our consideration of the decay of the radius of gyration in the investigation of the collapse time provides important information on the dynamics, but it is not sufficient to identify the different stages of the collapse of the initially coiled structures. 
Thus, to understand the nonequilibrium dynamics further, we decided to perform a study in line with 
the literature on standard phase-ordering kinetics, such as by determining how the 
average number of clusters evolves or by monitoring how the average size of the clusters grows with time. In the next subsection we present results from such analyses.
\subsection{Growth kinetics of clusters}
\par
Following the method of identifying the clusters along the polymer described in the previous subsection, we have calculated for each time evolution the mean cluster size $C_s$, as 
\begin{equation}
	C_s(t) = \frac{1}{n_c(t)} \sum_{k=1}^{n_c(t)} m_k,
\end{equation}
where $n_c(t)$ is the number of clusters in a chain at time $t$ and $m_k$ 
is the mass of the $k$-th cluster, which  measures the number of monomers 
within that cluster. 
\begin{figure}[t!]
	\centering
	\includegraphics*[width=0.45\textwidth]{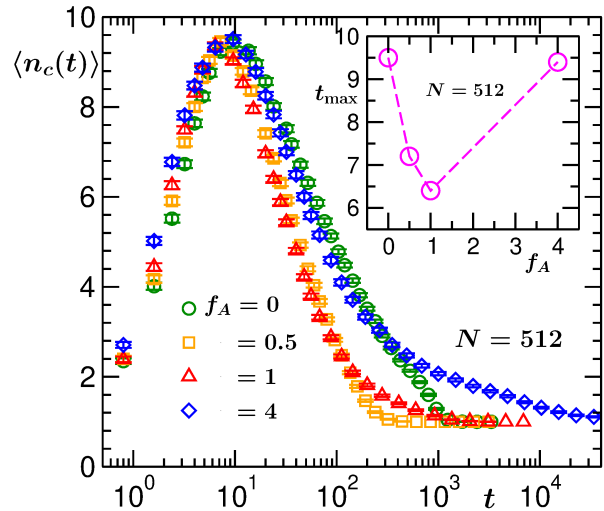}
	\caption{\label{nc_t} Semi-log plots of the time evolution of the average number of clusters  $\langle n_c(t) \rangle$ for several
		$f_A$ values with $N=512$. The inset shows the variation of $t_{\rm{max}}$, 
		the time at which $\langle n_c(t) \rangle$ reaches its maximum, as a function of $f_A$.}
\end{figure}
\par 
In Fig.~\ref{nc_t} we plot the average number of clusters $\langle n_c(t) \rangle$ as a function of 
time on a semi-log scale for the passive model as well as for the active cases. During the initial period, we observe a rapid rise in $ \langle n_c \rangle$ for all values of $f_A$. This is reminiscent of the nucleation phenomena observed during the vapor-liquid transition in a system of particles \cite{roy_13}. 
From this plot it is hard to identify any minor differences in the values of  $t_{\rm{max}}$ 
at which $\langle n_c(t) \rangle$ reaches its maximum for different $f_A$. However, with a closer look one identifies a weak nonmonotonicity in $t_{\rm{max}}$ with the variation of $f_A$ which we show in the inset of Fig.~\ref{nc_t}. 
\par 
Past this peak, in the second stage, $ \langle n_c \rangle$ decreases with $t$, suggesting the beginning of coarsening. During this period, clusters merge with each other and form a bigger one. In this regime $f_A$ has significant impact on $\langle n_c \rangle$. There one observes a nonmonotonic behavior for the times at which the polymer becomes a globule, i.e., $\langle n_c \rangle=1$. For $f_A=0.5$, $ \langle n_c \rangle$ reaches unity much earlier than in the passive case. With further increase in $f_A$, the trend gets altered, consistent with the time evolutions depicted in Fig.~\ref{snap}. This also reconfirms our earlier observation from Fig.~\ref{rg_fa} that the time required for the full decay of $R_g^2$ (which is related to the formation of the globule) is much longer for strong activity with $f_A=4$. This may have its origin in the fact that once the velocity ordering is strong, it takes much longer for the last two clusters to come closer to each other and form a single globule.
\par
\begin{figure}[t!]
	\centering
	\includegraphics*[width=0.44\textwidth]{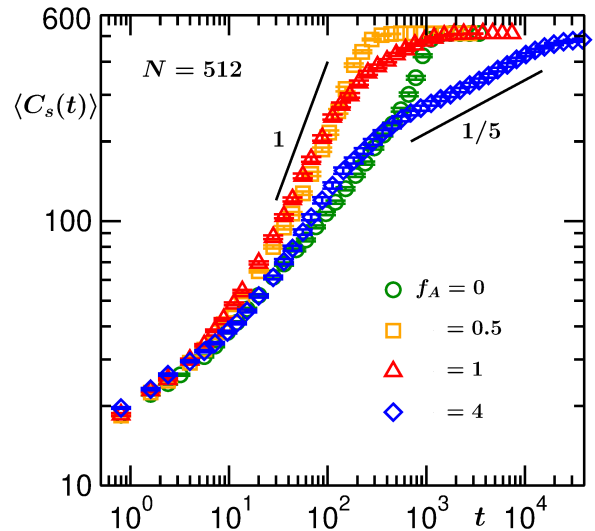}
	\caption{\label{clsize_t} Log-log plots of the average cluster size  $\langle C_s(t) \rangle$  
		versus time, for different values of $f_A$, with $N=512$. 
		The solid lines are drawn as guides to the eyes with power-law exponents $1$ and $1/5$.}
\end{figure}

 
In Fig.~\ref{clsize_t} we plot the average size of the clusters, $\langle C_s \rangle$,
as a function of time on a log-log scale for all four $f_A$ values. 
In general, it follows a power-law behavior,
\begin{equation}
\langle C_s(t) \rangle \sim t^{\alpha_c}\,,
\end{equation}
where $\alpha_c$ is the cluster-growth exponent. After an initial transient period the coarsening scaling regime starts where
small clusters merge with each other  and form larger ones. First we discuss the passive case. 
For this, in the scaling regime, the growth exponent is $\alpha_c \approx 1$, consistent with earlier results from Monte Carlo simulations \cite{majumder1, majumder3}. However, from our MD 
simulations using Langevin equations, $\alpha_c$ appears to be somewhat smaller than $1$. Due to continuous bending of the data it is difficult, however, to 
fit a suitable power law over a longer period.  We hence put  a line with the exponent $1$ as a guide to the eye.  The initial transient stage is present for the active cases as well.  Later, in the scaling regime, data for the active cases seem to follow different growth laws than in the passive case. For $f_A=0.5$ the growth is faster with still $\alpha_c$ $\approx 1$ in the scaling regime.
Data for $f_A=1$ follow a similar trend as that for the $f_A=0.5$ case up to $t \approx 200$, i.e., until the formation of a two-cluster conformation. Then the growth becomes slower for which the exponent appears to be $\approx 1/5$, until a single globule forms and the finite-size limit is reached.  Now, with $f_A=4$ also, at initial times until $t \approx 10$ (close to the value of $t_{\rm{max}}$, cf.\ Fig.~\ref{nc_t}) data follow a similar trend as for $f_A=0.5$ and $1$. Then, in the scaling regime, which is quite prolonged in this case, the growth becomes much slower compared to the lower activities and thus  $\alpha_c$ is significantly smaller than $1$. 
In this regime data look consistent with $\alpha_c \approx 1/5$, similar to the $f_A=1$ case, before the finite-size limit is reached.
This slower growth over an extended period indicates a longer persistence of the dumbbell conformations which was also observed from the snapshots presented in Fig.~\ref{snap}. In this period the clusters grow in size only by taking beads from the bridges connecting them rather than via merging of the clusters. 
\par
\begin{figure}[t!]
	\centering
	\includegraphics*[width=0.45\textwidth, angle=0]{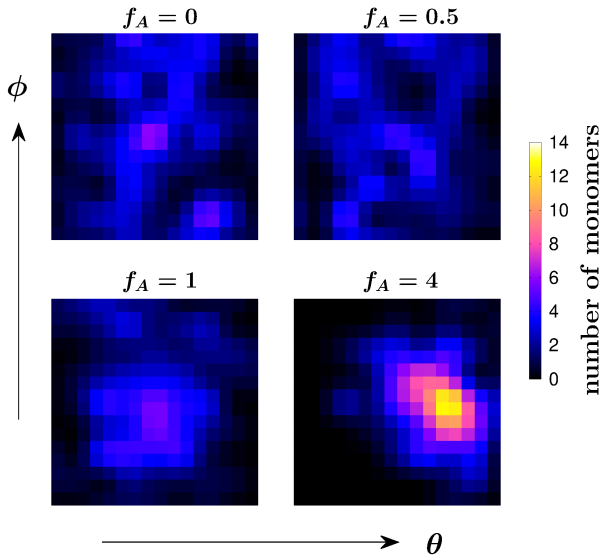}
	\caption{\label{angle_distri} Plots of the distributions of the velocity 
orientations of the beads, 
		depicted via the angles $\theta$ and $\phi$, for the largest cluster present along the polymer chain at the time when $n_c(t)$ is the maximum for a typical run, 
		for all the considered values of $f_A$. The ranges for $\theta$ and $\phi$ in each 
plot are $[0,\pi]$ and $[-\pi,\pi]$, respectively.}
\end{figure}

To explain the above mentioned nonmonotonicity in the coarsening we looked at the velocity orientations of the beads for a typical time evolution run. First, we choose the conformations for which 
the number of clusters along the polymer is maximum for various choices of $f_A$ values. 
The corresponding times are quoted in the inset of Fig.~\ref{nc_t}. We  then recorded the orientational ordering of the beads within the largest cluster. For this, we calculated two angles, namely the polar and the azimuthal angles (in spherical coordinates) for the velocities of the beads, denoted by 
$\theta$ and $\phi$, respectively, defined as
\begin{eqnarray}\label{angles}
\theta_j ={\rm{cos}}^{-1} \Bigg(\frac{v_{j,z}}{|\vec{v}_j|}\Bigg) ~{\rm{and} }
~~\phi_j ={\rm{tan}}^{-1} \Bigg(\frac{v_{j,y}}{v_{j,x}}\Bigg)
\end{eqnarray}
with  $\vec{v}_j~ [=(v_{j,x},v_{j,y},v_{j,z})]$ denoting the velocity of the $j$-th bead.
In Fig.~\ref{angle_distri} we plot their distributions for all the studied values of $f_A$. 
It is clearly seen that for $f_A=0$ data are uniformly distributed
over the whole range of $\theta$ and $\phi$.  With the increase of $f_A$,
we see that the distribution becomes confined within a certain region of $\theta$ and $\phi$. This suggests
ordering of velocities of the beads in a particular direction within a typical cluster. Though we present results for the largest cluster, we checked that this fact is true for the other clusters along the chain as well. This indicates that for higher activity the ordering of velocity occurs faster than the coarsening in density field. To understand the nonmonotonic behavior of the cluster growth as a function of $f_A$,
it is instructive to monitor the orientation of the  center-of-mass velocities of different clusters along the chains. More specifically, here we want to check whether the alignment interaction, which is applied locally for each bead, affects all the clusters along 
the chain. It is possible that the alignment propagates even through the non-clustered regions of the polymer.
\par
\begin{figure}[t!]
	\centering
	\includegraphics*[width=0.44\textwidth]{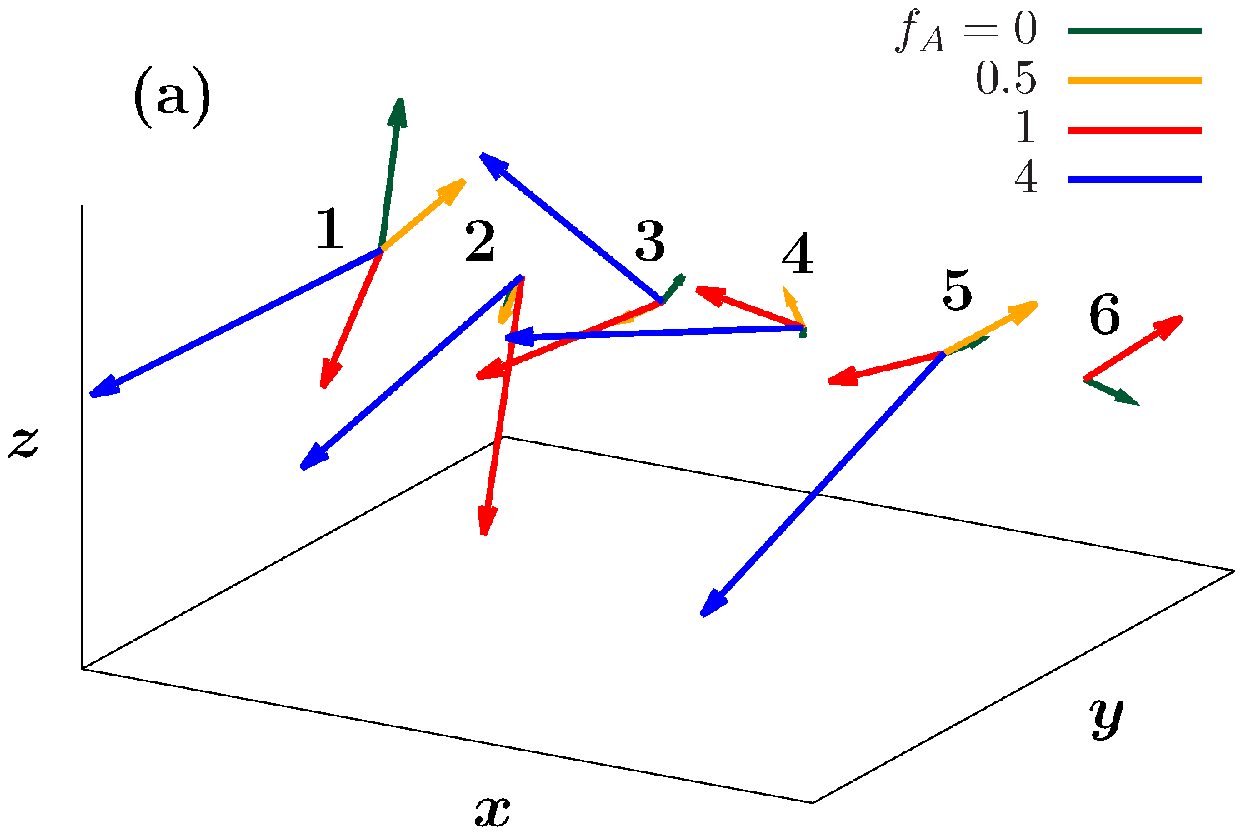}
	\vskip 0.4cm
	\includegraphics*[width=0.41\textwidth]{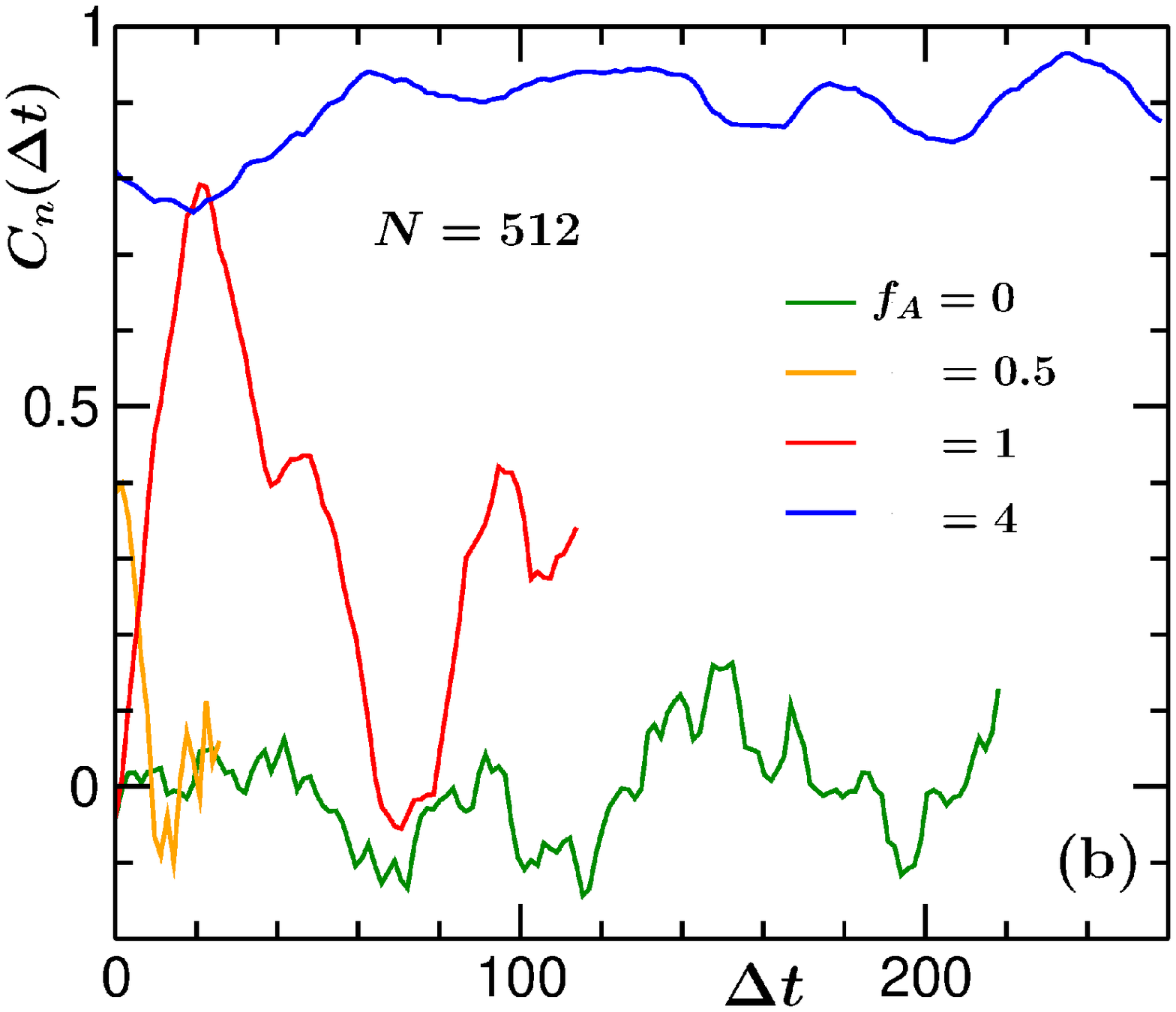}
	\caption{\label{com_velo}(a) Center-of-mass velocity vectors of clusters formed along a polymer chain in the
			scaling regime of coarsening for a typical time evolution with $N=512$.
		The serial numbers of the clusters are marked next to them. More technical details are mentioned in the text. (b) Plot of $C_n(\Delta t)$ versus $\Delta  t$ in the scaling regime for different values of $f_A$.}
\end{figure}

For the above purpose, we consider times in the scaling regime of the coarsening stage at which the polymer contains about $5$ large clusters  implying that the clusters are not very far apart from each other. For each of these clusters we calculate its center-of-mass velocity as
\begin{equation}
	\vec{v}_{k}^{\,\rm{cm}} = \frac{1}{m_k}\sum_{i=1}^{m_k}\vec{v}_i\,,
\end{equation}
where $k$ varies from $1$ to $n_c$ and $m_k$ is the number of beads in the $k$-th cluster. This is shown in Fig.~\ref{com_velo}(a) for all the considered $f_A$ values.  The vectors in this figure denote the velocities of the individual clusters, whose serial numbers $k=1, 2, 3, ...$ along the contour of the chain are marked next to them and different colors encode different values of $f_A$. For the clarity of understanding the center-of-mass positions of the clusters are artificially shifted to ($k,0,0$).  We see that for $f_A=0$ the directions of 
the center-of-mass velocities of the clusters are random along the whole chain and also their magnitudes are smaller. This is consistent with the saturation of $v_a$ close to $0$ in Fig.~\ref{opm} for the passive case. As an additional remark, we mention here that even after the formation of the globule the velocities of the beads remain random and this suggests the diffusive motion of the chain for $f_A=0$ \cite{paul_soft20}.
For $f_A=0.5$ we see that though the magnitudes are higher than  in the passive case, their directions are still quite random. 
This randomness helps the clusters to 
meet each other and larger velocities make them coalesce faster than in the passive case. Thus 
the coarsening happens faster for the $f_A=0.5$ case. For the higher values of $f_A$, however, as time progresses, the randomness in their velocity directions diminishes and
the local orientational ordering, which is expected due to the Vicsek-like  alignment interaction, gradually affects all the beads along the whole chain. This feature should be more prominent at later times.
\par 
For a quantitative understanding we considered the typical conformations of the polymer in the scaling regime. We measure how the correlation builds up among the directions of velocities of the center-of-mass for different clusters along the chain. Such a correlation for a typical run can be calculated as
	\begin{equation}\label{dir_correl}
	C_n(\Delta t)=\langle \hat{n}_j^{\,\rm{cm}}(t) \cdot  ~\hat{n}_k^{\,\rm{cm}}(t) \rangle \,,
	\end{equation}
	where $\langle ... \rangle$ in this case represents the averaging over different combinations of clusters for that run and $\hat{n}_j^{\rm{cm}}$ denotes the direction of velocity of the $j$-th cluster,  defined as 
	\begin{equation}
	 \hat{n}_j^{\,\rm{cm}} = \frac{\vec{v}_j^{\,\rm{cm}}}{|\vec{v}_j^{\,\rm{cm}}|}.	
	\end{equation}
	In Eq.~\eqref{dir_correl} $\Delta t = t - t_0$ represents the time span over which $C_n$ is calculated.
	For a given time evolution, we start the measurements at time $t_0$ when, after the initial coarsening, $n_c = 3$ clusters along the chain are present and stop them when any two clusters merge and $n_c$ becomes 2. The results are shown in Fig.~\ref{com_velo}(b). Note that
	the resulting time intervals for different $f_A$ values cannot be easily related to Fig.~\ref{nc_t} which shows averaged data.
 For $f_A=0$ we see that $C_n(\Delta t)$ remains always close to $0$. This is due to the randomness in the directions of different clusters. The value of $C_n(\Delta t)$ increases with the increase of $f_A$. For $f_A=4$ it becomes $\approx 0.9$ indicating that a particular direction of orientation is preferred by all  the clusters along the chain and it takes a longer time for two clusters to merge with each other, thus quantifying the qualitative picture of Fig.~\ref{snap}.  For higher activities, such a rapid velocity ordering of all the beads in different clusters along the whole chain occurs much earlier and also faster which eventually makes the coarsening process slower.
\section{ Conclusion}\label{conclusion}
We have studied the nonequilibrium dynamics of the coil-globule transition for a flexible homopolymer chain consisting of active beads. 
The quench temperature is chosen well below the collapse transition temperature, known for the passive polymer, 
such that one expects globular phases for the active cases as well. To study its kinetics, we have used MD simulation with Langevin thermostat 
to ensure that the temperature remains fixed at our choice.
The activity in our study has been incorporated in Vicsek-like manner which biases the beads to align 
their velocities in a direction decided by its neighboring beads. Thus, there exists a 
competition between the thermal fluctuations and the activity, and this fact makes the pathways of globule formation much interesting.
\par 
The primary aim of this paper is to investigate the effects of  activity on the relaxation or 
collapse time for the coil-globule transition of the polymer chain. Unexpectedly, this turns out to be very compelling as we observe a nonmonotonic behavior with increasing activity. 
Furthermore, for higher activity the distribution of the relaxation times,  $\tau_{\rm{cl}}$, calculated from different production runs becomes highly non-Gaussian, developing a long thin tail towards large $\tau_{\rm{cl}}$.  Eventually, for $f_A=4$ the peak for small $\tau_{\rm{cl}}$ vanishes and the distribution exhibits a crossover to a purely exponentially decaying behavior. The scaling of the average relaxation times with respect to the size of the polymer chain is also studied for different $f_A$ values.
The scaling exponent $z$ for $\tau_{\rm{cl}}^{\rm{av}}$ reflects again the nonmonotonic behavior: For low activity it is smaller and for high activity larger than in the passive case. The growth kinetics and the evolution of the
number of clusters during different stages of collapse have been identified and these results are in good agreement with the aforementioned facts related to the relaxation times. 
The cluster-growth exponent for the lower activities is compatible with $\alpha_c \approx 1$ as in the passive case, but for $f_A=4$ it appears to be significantly smaller with $\alpha_c \approx 1/5$. 
 
\par
We have performed the simulations using the Langevin thermostat which takes into account the solvent properties only implicitly. As a future project it would be interesting to model the solvent properties more faithfully in the presence of activity, in particular by including the hydrodynamic effects.  Also experimental realizations of an active polymer using colloidal or Janus particles can be interesting \cite{ramirez,daiki}. For this  one needs to find a suitable way with which the beads can be made active and also the degree of alignment can be modulated by tuning the activity strength. Studies with such a setup may be insightful for exploring the conformational dynamics of active polymers with varying activity.
\par
Beside these more conceptual aspects, it would be also important to quantify the effects of quench temperature
on the kinetics of coil-globule transitions. Such studies do exist for the case of passive polymers \cite{majumder3}, 
for which it has been shown that there exists a master curve for cluster growth. This kind of study will 
be interesting for active polymers also, with the objective to search for 
universal features.\\
\section*{Acknowledgements}
This project was funded by the Deutsche Forschungsgemeinschaft (DFG, German Research Foundation) 
under Grant No.\ 189\,853\,844--SFB/TRR 102 (Project B04). It was further 
supported by the Deutsch-Franz\"osische Hochschule (DFH-UFA) 
through the Doctoral College ``$\mathbb{L}^4$'' under Grant No.\ CDFA-02-07, the Leipzig Graduate School of Natural Sciences ``BuildMoNa'', 
and the EU COST programme EUTOPIA under Grant No.\ CA17139.
\end{document}